\documentclass[prl,twocolumn,superscriptaddress,showpacs,nofootinbib]{revtex4-1}
\usepackage{graphicx}
\usepackage{amssymb}
\usepackage{mathrsfs}
\usepackage{amsmath}
\usepackage{color}

\allowdisplaybreaks[1]

\begin{document}

\title{Strong CP Problem, Neutrino Masses and the 750 GeV Diphoton Resonance}

\author{Qing-Hong Cao}
\email{qinghongcao@pku.edu.cn}
\affiliation{Department of Physics and State Key Laboratory of Nuclear Physics and Technology, Peking University, Beijing 100871, China}
\affiliation{Collaborative Innovation Center of Quantum Matter, Beijing 100871, China}
\affiliation{Center for High Energy Physics, Peking University, Beijing 100871, China}

\author{Shao-Long Chen}
\email{chensl@mail.ccnu.edu.cn}
\affiliation{Key Laboratory of Quark and Lepton Physics (MoE) and Institute of Particle Physics, Central China Normal University, Wuhan 430079, China}
\affiliation{Center for High Energy Physics, Peking University, Beijing 100871, China}

\author{Pei-Hong Gu}
\email{peihong.gu@sjtu.edu.cn}
\affiliation{Department of Physics and Astronomy, Shanghai Jiao Tong University, 800 Dongchuan Road, Shanghai 200240, China}

\begin{abstract}
We present an $SU(3)^{}_{c}\times SU(2)^{}_{L}\times SU(2)^{}_{R}\times U(1)_{L}^{}\times U(1)_{R}^{}\rightarrow SU(3)^{}_{c}\times SU(2)^{}_{L}\times SU(2)^{}_{R}\times U(1)^{}_{B-L}$ left-right symmetric model with a discrete parity symmetry to realize a universal seesaw scenario. The model can simultaneously solve the strong CP problem without resorting to the unobserved axion and explain the 750 GeV diphoton resonance reported recently by the ATLAS and CMS collaborations at the LHC. Owing to large suppressions in the two-loop induced Dirac mass terms, the Majorana mass matrices of left- and right-handed neutrinos naturally share the same structure. That allows us to quantitatively study the neutrinoless double beta decay induced by the right-handed currents.
\end{abstract}

\pacs{14.60.Pq, 12.60.Cn, 12.60.Fr}

\maketitle

\section{I. Introduction}

The discovery of the 125 GeV Higgs boson at the LHC~\cite{Aad:2012tfa,Chatrchyan:2012xdj} is a landmark which indicates
the underlying reason of the spontaneous electroweak symmetry breaking and the mass generation mechanism. One
question often raised is, in additon to the Higgs boson, whether or not there are other fundamental particles or mechanisms also play roles in the electroweak symmetry breaking. Especially, the smallness of neutrino masses is a puzzle which drives us to go beyond the
standard model (SM). The seesaw mechanism~\cite{minkowski1977,yanagida1979,grs1979,ms1980} provides a natural way to
understand the tiny size of the observed neutrino masses. Generalized seesaw mechanism (called universal seesaw)
~\cite{Berezhiani:1983hm,Chang:1986bp,Rajpoot:1987fca,Davidson:1987mh}
were introduced to generate quark and charged lepton masses under a ``seesaw'' way, with an extension of a new
set of vector-like singlet fermions (quarks and leptons). The universal seesaw can be elegantly achieved in some $SU(3)^{}_c\times SU(2)^{}_L\times SU(2)^{}_R\times U(1)^{}_{B-L}$ left-right symmetric models \cite{ps1974} where the spontaneous breaking of left-right and electroweak symmetries are driven by two $[SU(2)]$-doublet Higgs scalars, while the ordinary charged fermions from $SU(2)$ doublets obtain their masses by integrating out additional $[SU(2)]$-singlet
charged fermions. In such left-right symmetric models for the universal seesaw, one can further solve the strong CP problem without resorting to the unobserved axion if the discrete parity symmetry is imposed \cite{ms1978,bm1989}.

In this work, we shall consider an $SU(3)^{}_c\times SU(2)^{}_L\times SU(2)^{}_R\times U(1)^{}_{L}\times U(1)^{}_{R}$ left-right symmetric model where the $[SU(2)]$-singlet fermions for the universal seesaw can naturally acquire their masses after a spontaneous symmetry breaking $U(1)_{L}^{}\times U(1)_{R}^{} \rightarrow U(1)^{}_{B-L}$. For this symmetry breaking, we introduce some $[SU(2)]$-singlet Higgs scalars which provide a natural solution to the 750 GeV diphoton resonance observed recently at the LHC Run-2 with a center-of-mass energy $\sqrt{s}=13$ TeV~\cite{atlas, cms}. It has drawn lots of interests in the field~\cite{Backovic:2015fnp,Harigaya:2015ezk,Mambrini:2015wyu,Angelescu:2015uiz,Pilaftsis:2015ycr,Franceschini:2015kwy,DiChiara:2015vdm,Low:2015qep,Bellazzini:2015nxw,Ellis:2015oso,McDermott:2015sck,Higaki:2015jag,Gupta:2015zzs,Petersson:2015mkr,Molinaro:2015cwg,Nakai:2015ptz,Buttazzo:2015txu,Bai:2015nbs,Aloni:2015mxa,Falkowski:2015swt,Csaki:2015vek,Agrawal:2015dbf,Ahmed:2015uqt,Chakrabortty:2015hff,Bian:2015kjt,Curtin:2015jcv,Fichet:2015vvy,Chao:2015ttq,Demidov:2015zqn,No:2015bsn,Becirevic:2015fmu,Cox:2015ckc,Kobakhidze:2015ldh,Matsuzaki:2015che,Cao:2015pto,Dutta:2015wqh,Benbrik:2015fyz,Megias:2015ory,Carpenter:2015ucu,Bernon:2015abk,Alves:2015jgx,Gabrielli:2015dhk,Chao:2015nsm,Arun:2015ubr,Han:2015cty,Chang:2015bzc,Chakraborty:2015jvs,Ding:2015rxx,Han:2015dlp,Han:2015qqj,Luo:2015yio,Chang:2015sdy,Bardhan:2015hcr,Feng:2015wil,Antipin:2015kgh,Wang:2015kuj,Cao:2015twy,Huang:2015evq,Liao:2015tow,Heckman:2015kqk,Bi:2015uqd,Cho:2015nxy,Cline:2015msi,Bauer:2015boy,Boucenna:2015pav, Murphy:2015kag, Hernandez:2015ywg,Dey:2015bur, Pelaggi:2015knk,deBlas:2015hlv, Belyaev:2015hgo, Dev:2015isx}.
We further add two $[SU(2)]$-triplet Higgs scalars to generate the Majorana mass matrices of the left- and right-handed
neutrinos. The Dirac mass term between the left- and right-handed neutrinos can be highly suppressed since it is produced only
by the charged gauge bosons and fermions at two-loop level. The right-handed neutrino mass matrix is proportional to the left-handed neutrino mass matrix so that the neutrinoless double beta decay from the right-handed currents can be quantitatively studied~\cite{bgmr2013}. Our model can also accommodate the parity symmetry for solving the strong CP problem without an axion.

The paper is organized as follows: In Sec. II and III we present our model and the pattern of the symmetry breaking.
In Sec. IV and V,  we address on the charged fermion mass and neutrino mass generation and how to solve the strong CP problem. In Sec. VI we interpret the 750 GeV diphoton resonance as the singlet scalar, the mass generator of the vectorquark in our model. Finally, we conclude in Sec. VII.

\section{II. The model}

In the fermion sector we have the following $SU(2)$ singlets,
\begin{subequations}
\begin{eqnarray}
\begin{array}{l}U^{}_{L}(3,1,1,-\frac{4}{3},0)\end{array}
&\stackrel{P}
{\longleftrightarrow}& \begin{array}{l}U^{}_{R}(3,1,1,0,-\frac{4}{3})\end{array},\\
[2mm]
\begin{array}{l}D^{}_{L}(3,1,1,+\frac{2}{3},0)\end{array}&\stackrel{P}
{\longleftrightarrow}& \begin{array}{l}D^{}_{R}(3,1,1,0,+\frac{2}{3})\end{array},\\
[2mm]
\begin{array}{l}E^{}_{L}(1,1,1,+2,0)\end{array}&\stackrel{P}
{\longleftrightarrow}& \begin{array}{l}E^{}_{R}(1,1,1,0,+2)\end{array},
\end{eqnarray}
\end{subequations}
in addition to the $SU(2)$ doublets as below,
\begin{subequations}
\begin{eqnarray}
&&\begin{array}{c}
q^{}_L(3,2,1,+\frac{1}{3},0)\\
[2mm]
\uparrow\\
\!\!\!\!\!\!P\,|\\
\downarrow\\
[2mm]
q^{}_R(3,1,2,0,+\frac{1}{3})
\end{array}\begin{array}{l}=\left[\begin{array}{c}u_L^{}\\
[2mm]
d_L^{}\end{array}\right]\\
[14mm]
=\left[\begin{array}{c}u_R^{}\\
[2mm]
d_R^{}\end{array}\right],\end{array}\\
[5mm]
&&\begin{array}{c}
l^{}_L(1,2,1,-1,0)\\
[2mm]
\uparrow\\
\!\!\!\!\!\!P\,|\\
\downarrow\\
[2mm]
l^{}_R(1,1,2,0,-1)
\end{array}\begin{array}{l}=\left[\begin{array}{c}\nu_L^{}\\
[2mm]
e_L^{}\end{array}\right]\\
[14mm]
=\left[\begin{array}{c}N_R^{}\\
[2mm]
e_R^{}\end{array}\right].\end{array}
\end{eqnarray}
\end{subequations}
Here and thereafter the brackets following the fields describe the transformations under the $SU(3)^{}_c\times SU(2)^{}_L\times SU(2)^{}_R\times U(1)^{}_{L}\times U(1)^{}_{R}$ gauge groups. As for the Higgs scalars, they include
three $SU(2)$ singlets,
\begin{subequations}
\begin{eqnarray}
\begin{array}{l}
\sigma^{}_U(1,1,1,+\frac{4}{3},-\frac{4}{3})\end{array}&\stackrel{P}
{\longleftrightarrow}&\sigma^\ast_U,\\
[2mm]
\begin{array}{l}
\sigma^{}_D(1,1,1,-\frac{2}{3},+\frac{2}{3})\end{array}&\stackrel{P}
{\longleftrightarrow}&\sigma^\ast_D,\\
[2mm]
\begin{array}{l}
\sigma^{}_E(1,1,1,-2,+2)\end{array}&\stackrel{P}
{\longleftrightarrow}&\sigma^\ast_E,
\end{eqnarray}
\end{subequations}
two $SU(2)$ doublets,
\begin{eqnarray}
\begin{array}{c}
\phi^{}_L(1,2,1,-1,0)\\
[2mm]
\uparrow\\
\!\!\!\!\!\!P\,|\\
\downarrow\\
[2mm]
\phi^{}_R(1,1,2,0,-1)
\end{array}\begin{array}{l}=\left[\begin{array}{c}\phi_L^{0}\\
[2mm]
\phi_L^{-}\end{array}\right]\\
[12mm]
=\left[\begin{array}{c}\phi_R^{0}\\
[2mm]
\phi_R^{-}\end{array}\right],\end{array}
\end{eqnarray}
as well as two $SU(2)$ triplets,
\begin{eqnarray}
&&\begin{array}{c}
\Delta^{}_L(1,3,1,+2,0)\\
[2mm]\\
\uparrow\\
\!\!\!\!\!\!P\,|\\
\downarrow\\
[2mm]
\Delta^{}_R(1,1,3,0,+2)\end{array}
\begin{array}{l}=\left[\begin{array}{cc}\frac{1}{\sqrt{2}}\delta_L^{+}&\delta^{++}_{L}\\
[5mm]
\delta^{0}_{L}&-\frac{1}{\sqrt{2}}\delta^{+}_L\end{array}\right]\\
[16mm]
=\left[\begin{array}{cc}\frac{1}{\sqrt{2}}\delta_R^{+}&\delta^{++}_{R}\\
[5mm]
\delta^{0}_{R}&-\frac{1}{\sqrt{2}}\delta^{+}_R\end{array}\right].\end{array}
\end{eqnarray}
The allowed parity-invariant Yukawa interactions then should be
\begin{eqnarray}
\label{yukawa}
\mathcal{L}_{Y}^{}&=&-y^{}_{U}(\bar{q}^{}_L\phi^{}_L U^{c}_L+\bar{q}^{}_R\phi^{}_R U^{c}_R)-f^{}_U\sigma_{U}^{}\overline{U}^{}_L U^{}_R\nonumber\\
&&-y^{}_{D}(\bar{q}^{}_L\tilde{\phi}^{}_L D^{c}_L+\bar{q}^{}_R\tilde{\phi}^{}_R D^{c}_R)-f^{}_D\sigma_{D}^{}\overline{D}^{}_L D^{}_R\nonumber\\
&&-y^{}_{E}(\bar{l}^{}_L\tilde{\phi}^{}_L E^{c}_L+\bar{l}^{}_R\tilde{\phi}^{}_R E^{c}_R)-f^{}_E\sigma_{E}^{}\overline{E}^{}_L E^{}_R\nonumber\\
&&-\frac{1}{2}f_\Delta^{}(\bar{l}^c_L i\tau^{}_2\Delta_L^{} l_L^{}+\bar{l}^c_R i\tau^{}_2\Delta_R^{} l_R^{}) +\textrm{H.c.}
~~\textrm{with}\nonumber\\
&&f_{U,D,E}^{}=f_{U,D,E}^{\dagger}\,,~~f_\Delta^{}=f_\Delta^T\,.
\end{eqnarray}
For simplicity, we shall not write down the full scalar potential where the parity symmetry is softly broken. Instead, we only give the terms relevant to our demonstration,
\begin{eqnarray}
\label{potential}
V&\supset&\lambda(\sigma_E^{\ast}\sigma_D^{3}+\textrm{H.c.})+\xi(\sigma_U^{}\sigma_D^2+\textrm{H.c.})\nonumber\\
&&+\eta(\sigma_E^{}\sigma_D^{}\sigma^\ast_U
+\textrm{H.c.})+\mu^{2}_{\phi^{}_{L}}\phi^\dagger_{L}\phi^{}_{L}+\mu^2_{\phi_R^{}}\phi^\dagger_R\phi^{}_R\nonumber\\
&&
+\mu^{2}_{\Delta^{}_{L}}\textrm{Tr}(\Delta^\dagger_{L}\Delta^{}_{L})
+\mu^2_{\Delta^{}_R}\textrm{Tr}(\Delta^\dagger_R\Delta^{}_R)\nonumber\\
&&
+\rho_L^{}(\phi^T_{L}i\tau_2^{}\Delta_{L}^{}\phi_{L}^{}+\textrm{H.c.})+\rho_R^{}(\phi^T_{R}i\tau_2^{}\Delta_R^{}\phi_R^{}+\textrm{H.c.})\nonumber\\
&&\textrm{with}~~\mu^{2}_{\phi^{}_{L}}\neq \mu^{2}_{\phi^{}_{R}}\,,
~~\mu^{2}_{\Delta^{}_{L}}\neq \mu^{2}_{\Delta^{}_{R}}\,,~~\rho_L^{}\neq \rho_R^{}\,.
\end{eqnarray}

\section{III. Symmetry breaking}

The $[SU(2)]$-singlet Higgs scalar $\sigma_D^{}$ is responsible for spontaneously breaking the $U(1)_{L}^{}\times U(1)_{R}^{}$ symmetry down to a $U(1)_{B-L}^{}$ symmetry, i.e.
\begin{eqnarray}
\label{ssbbl1}
U(1)^{}_L\times U(1)^{}_{R} \stackrel{\langle\sigma^{}_{D}\rangle}{\longrightarrow}  U(1)^{}_{B-L}~~\textrm{with}~~ \langle\sigma^{}_{D}\rangle=\frac{v_D^{}}{\sqrt{2}}\,.
\end{eqnarray}
The other $[SU(2)]$-singlet Higgs scalars $\sigma_{U,E}^{}$ then will acquire their vacuum expectation values (VEVs),
\begin{eqnarray}
\label{ssbbl2}
\langle\sigma_{U}^{}\rangle&=&\frac{v_U^{}}{\sqrt{2}}\simeq -\frac{\xi\langle\sigma^{}_{D}\rangle^2_{}}{M_{\sigma_U^{}}^2}\,,\\
\langle\sigma_{E}^{}\rangle&=&\frac{v_E^{}}{\sqrt{2}}\simeq -\frac{\eta\langle\sigma^{}_{D}\rangle\langle\sigma^{}_{U}\rangle
+\lambda\langle\sigma^{}_{D}\rangle\langle\sigma^{2}_{U}\rangle}{M_{\sigma_E^{}}^2}\,.
\end{eqnarray}
We can rewrite the $[SU(2)]$-singlet Higgs scalars as
\begin{subequations}
\begin{eqnarray}
\sigma_D^{}&=&\frac{1}{\sqrt{2}}(v_D^{}+H_D^{}+iA_D^{})\,,\\
\sigma_U^{}&=&\frac{1}{\sqrt{2}}(v_U^{}+H_U^{}+iA_U^{})\,,\\
\sigma_E^{}&=&\frac{1}{\sqrt{2}}(v_E^{}+H_E^{}+iA_E^{})\,.
\end{eqnarray}
\end{subequations}
The Higgs bosons $H_{D,U,E}^{}$ will mix with each other. One of the linear combinations of the Goldstone bosons $A_{D,U,E}^{}$ will be eaten by the gauge boson corresponding to the $U(1)_{L}^{}\times U(1)_{R}^{}\rightarrow U(1)_{B-L}^{}$ symmetry breaking.

When the $[SU(2)_R^{}]$-doublet Higgs scalar $\phi^{}_{R}$ develops its VEV,
\begin{eqnarray}
\label{ssblr1}
\langle\phi^{}_R\rangle=\left[\begin{array}{c}\dfrac{v_R^{}}{\sqrt{2}}\\
[2mm]
0\end{array}\right]\,,
\end{eqnarray}
the left-right symmetry will be spontaneously broken down to the electroweak symmetry, i.e.
\begin{eqnarray}
\label{ssblr2}
&&SU(3)^{}_c\times SU(2)^{}_L\times SU(2)^{}_R\times U(1)^{}_{B-L} \nonumber\\
&& \stackrel{\langle\phi^{}_{R}\rangle}{\longrightarrow} SU(3)^{}_c\times SU(2)^{}_L\times U(1)^{}_Y \,.
\end{eqnarray}
Subsequently, the $[SU(2)_L^{}]$-doublet Higgs scalar $\phi^{}_{L}$ will drive the spontaneous electroweak symmetry breaking, i.e.
\begin{eqnarray}
\label{ssbew}
&&SU(3)^{}_c\times SU(2)^{}_L\times U(1)^{}_{Y} \stackrel{\langle\phi^{}_{L}\rangle}{\longrightarrow} SU(3)^{}_c\times U(1)^{}_{em} \nonumber\\
[2mm]
&&\quad\textrm{with}~~\langle\phi^{}_L\rangle=\left[\begin{array}{c}\dfrac{v_L^{}}{\sqrt{2}}\\
[2mm]
0\end{array}\right]\,.
\end{eqnarray}
Furthermore, the $[SU(2)]$-triplet Higgs scalars $\Delta_{L,R}^{}$ will acquire the induced VEVs,
\begin{eqnarray}
\langle\Delta_{L,R}^{}\rangle=\left[\begin{array}{cc}0&0\\
[5mm]
\frac{u^{}_{L,R}}{\sqrt{2}}&0\end{array}\right]~~\textrm{with}~~u_{L,R}\simeq -\frac{\rho_{L,R}^{}v^{2}_{L,R}}{\sqrt{2}M_{\delta_{L,R}^0}^2}\,.\nonumber\\
~
\end{eqnarray}
Obviously, the VEVs $\langle\sigma^{}_{D,U,E}\rangle$ and $\langle\phi^{}_{L,R}\rangle$ are all real. Since the parity symmetry has been softly broken in the scalar potential (\ref{potential}), the left-right and electroweak symmetry breaking can occur at different scales, i.e.
\begin{eqnarray}
\label{ssblrew}
\langle\phi_{R}^{}\rangle\gg \langle\phi_{L}^{}\rangle\,.
\end{eqnarray}

\section{IV. Charged fermion masses and solution to the strong CP problem}

\begin{figure}
\includegraphics[width=0.25\textwidth]{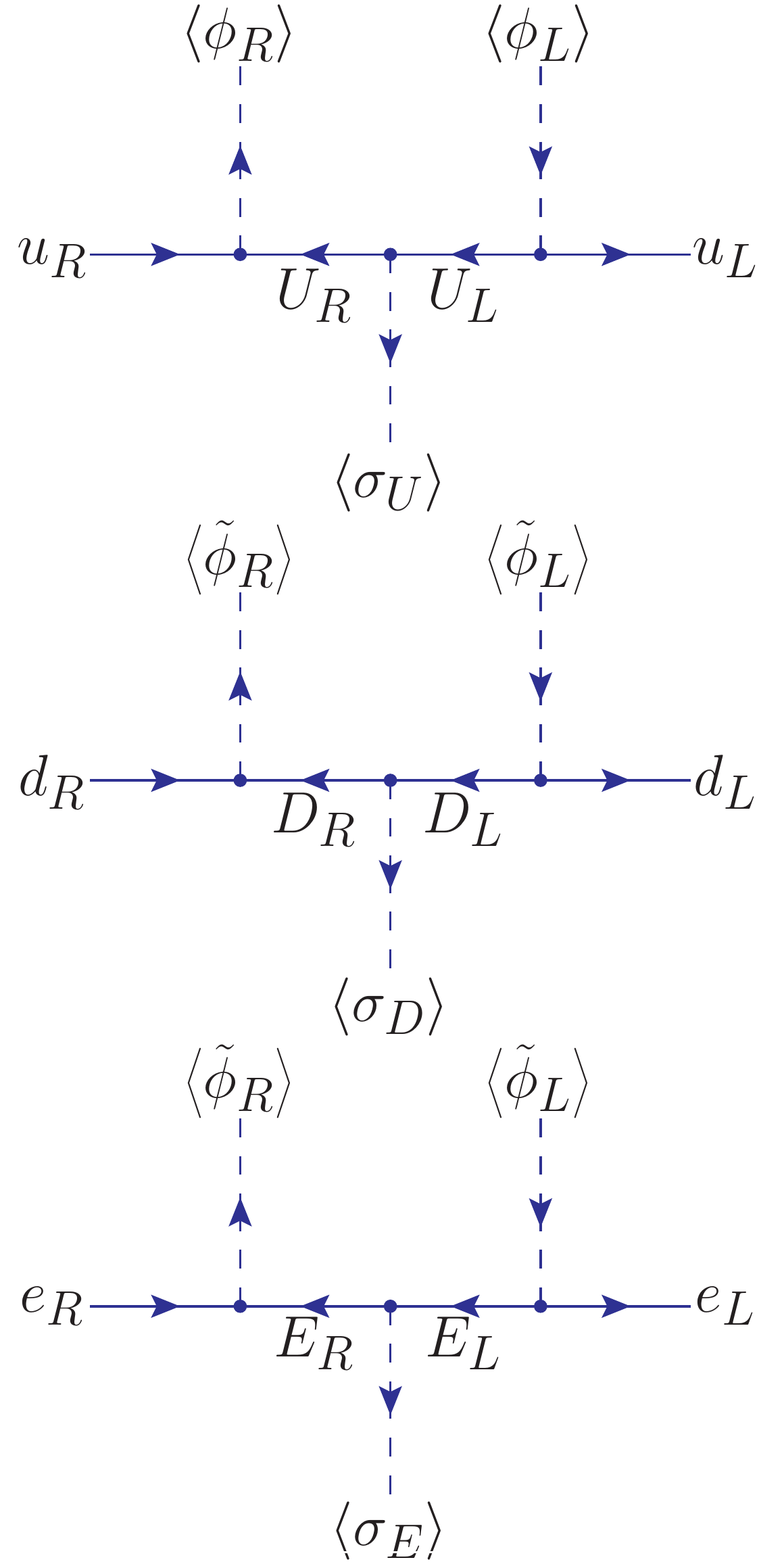}
\caption{\label{fmass}
Tree-level diagrams for generating the masses of up-type quarks (up), down-type quarks (middle) and charged leptons (down).}
\end{figure}

After the spontaneous symmetry breaking (\ref{ssbbl1}-\ref{ssblrew}), we can easily find the mass terms of the charged fermions,
\begin{eqnarray}
\label{qmass}
\!\!\!\!\mathcal{L}\!\!&\supset&\!\!-\left[\begin{array}{cc}\overline{u}^{}_L&\overline{U}^{c}_R\end{array}\right]
\left[\begin{array}{cc}0&\frac{y^{}_U v^{}_L}{\sqrt{2}}\\
[2mm]
\frac{y^\dagger_U v^{}_R}{\sqrt{2}}&\frac{f_U^{T} v^{}_E}{\sqrt{2}}\end{array}\right]\left[\begin{array}{c}u^{}_R\\
[2mm]
U^{c}_L\end{array}\right]\nonumber\\
[2mm]
&&\!\!-\left[\begin{array}{cc}\overline{d}^{}_L&\overline{D}^{c}_R\end{array}\right]
\left[\begin{array}{cc}0&\frac{y^{}_D v^{}_L}{\sqrt{2}}\\
[2mm]
\frac{y^\dagger_D v^{}_R}{\sqrt{2}}&\frac{f_D^{T} v^{}_D}{\sqrt{2}}\end{array}\right]\left[\begin{array}{c}d^{}_R\\
[2mm]
D^{c}_L\end{array}\right]\nonumber\\
[2mm]
&&\!\!-\left[\begin{array}{cc}\overline{e}^{}_L&\overline{E}^{c}_R\end{array}\right]
\left[\begin{array}{cc}0&\frac{y^{}_E v^{}_L}{\sqrt{2}}\\
[2mm]
\frac{y^\dagger_E v^{}_R}{\sqrt{2}}&\frac{f_E^{T} v^{}_E}{\sqrt{2}}\end{array}\right]\left[\begin{array}{c}e^{}_R\\
[2mm]
E^{c}_L\end{array}\right]+\textrm{H.c.}\,.
\end{eqnarray}
If the off-diagonal blocks in the above mass matrices are much lighter than the diagonal blocks, we can make use of the seesaw mechanism to give the masses of the usual charged fermions from the $SU(2)$ doublets,
\begin{equation}
\mathcal{L}\supset- \tilde{m}_u^{}\bar{u}^{}_L u^{}_R - \tilde{m}_d^{}\bar{d}^{}_L d^{}_R- \tilde{m}_e^{}\bar{e}^{}_L e^{}_R+\textrm{H.c.}~~
\end{equation}
with
\begin{eqnarray}
&&~\!\tilde{m}_{u}^{}=-y^{}_U\frac{v_L^{}v_R^{}}{2M_U^{}}y^\dagger_U=\tilde{m}_u^\dagger\nonumber\\
[1mm]
&&\quad~~=V^{}_u\hat{m}^{}_uV^{\dagger}_u=V^{}_u\textrm{diag}\{m_u^{},m_c^{},m_t^{}\}V^{\dagger}_u\,,\nonumber\\
[2mm]
&&~\!\tilde{m}_{d}^{}=-y^{}_D\frac{v_L^{}v_R^{}}{2M_D^{}}y^\dagger_D=\tilde{m}_d^\dagger\nonumber\\
[1mm]&&\quad~~=V^{}_d\hat{m}^{}_dV^{\dagger}_d
=V^{}_d\textrm{diag}\{m_d^{},m_s^{},m_b^{}\}V^{\dagger}_d\,,\nonumber\\
[2mm]
&&~\!\tilde{m}_{e}^{}=-y^{}_E\frac{v_L^{}v_R^{}}{2M_E^{}}y^\dagger_E=\tilde{m}_e^\dagger\nonumber\\
[1mm]&&\quad~~=U^{\dagger}_e \hat{m}^{}_e U^{}_e
=U^{\dagger}_e\textrm{diag}\{m_e^{},m_\mu^{},m_\tau^{}\}U^{}_e\,.
\end{eqnarray}
Here, $M_{U,D,E}^{}$ are the mass matrices of the additional charged fermions from the $SU(2)$ singlets,
\begin{eqnarray}
\mathcal{L}&\supset&- M_U^{}\overline{U}^{}_L U^{}_R - M_D^{}\overline{D}^{}_L D^{}_R- M_E^{}\overline{E}^{}_L E^{}_R+\textrm{H.c.}\nonumber
\end{eqnarray}
with
\begin{equation}
M_{U(D,E)}^{}=\frac{1}{\sqrt{2}}f^{}_{U(D,E)}v^{}_{U(D,E)}=M_{U(D,E)}^\dagger\,.\nonumber\\
\end{equation}
The seesaw scenario can be also understood from Fig. \ref{fmass}. Clearly, the CKM matrix should be $V=V_u^{}V^\dagger_d$ as usual. In the following, we will work in the base where the mass matrices $M_{U,D,E}^{}$ are real and diagonal.

Now the non-perturbative QCD Lagrangian should be
\begin{eqnarray}
\!\!\mathcal{L}_{QCD}^{}\supset\bar{\theta}\frac{g^2_3}{32\pi^2_{}}G\tilde{G}~~\textrm{with}~~
\bar{\theta}=\theta+\textrm{Arg}\textrm{Det} (M_u^{} M_d^{}),\nonumber\\
\end{eqnarray}
where $\theta$ is the phase from the QCD $\Theta$-vacuum while $M_u^{}$ and $M_d^{}$
are the mass matrices of the usual and additional down- and up-type quarks,
respectively,
\begin{eqnarray}
M_{d(u)}^{}=
\left[\begin{array}{cc}0&\dfrac{y_{D(U)}^{}v_L^{}}{\sqrt{2}}\\
[2mm]
\dfrac{y_{D(U)}^{\dagger}v_R^{}}{\sqrt{2}}&\dfrac{f_{D(U)}^{T}v_{D(U)}^{}}{\sqrt{2}}\end{array}\right]\,.
\end{eqnarray}
When the $\theta$-term is removed as a result of the
parity invariance, the real determinants
$\textrm{Det}(M_d^{})$ and $\textrm{Det}(M_u^{})$ will lead to a
zero $\textrm{Arg}\textrm{Det} (M_u^{} M_d^{})$. We hence obtain
a vanishing strong CP phase $\bar{\theta}$ at  tree level.

\section{V. Neutrino masses}
At tree level, through their Yukawa couplings with the $[SU(2)]$-triplet Higgs scalars, the left- and right-handed neutrinos will obtain their Majorana mass matrices
\begin{eqnarray}
\mathcal{L}&\supset&-\frac{1}{2}m_\nu^{\textrm{II}}\bar{\nu}_L^{}\nu_L^c -\frac{1}{2}M_N^{}\bar{N}_R^{}N_R^c+\textrm{H.c.}~~\textrm{with}\nonumber\\
&&m_\nu^{\textrm{II}}=f_\Delta^{}\langle\Delta_{L}^{}\rangle\,,~~M_N^{}=f_\Delta^{}\langle\Delta_{R}^{}\rangle\,.
\end{eqnarray}
The Dirac mass term between the left- and right-handed neutrinos is induced by a two-loop diagram mediated by the charged gauge bosons and fermions. One can roughly estimate
\begin{equation}
\mathcal{L}\supset -m_D^{}\bar{\nu}_L^{}N_R^{}+\textrm{H.c.}
\end{equation}
with
\begin{equation}
m_D^{}\sim \frac{3g^4_{}}{4(16\pi^2_{})^2}\frac{m_t^{}m_b^{}}{M_{W_R^{}}^2}\tilde{m}_e^{}\,.
\end{equation}
The full masses of the left-handed neutrinos then should be a sum of the type-I and II seesaw,
\begin{eqnarray}
m_\nu^{}=m_\nu^{\textrm{II}}-m_D^{}\frac{1}{M_N^{}}m_D^T\,.
\end{eqnarray}

It is easy to check that $m_D^{}\lesssim 6\,\textrm{eV}$ for $M_{W_R^{}}^{}\gtrsim1\,\textrm{TeV}$ and then $m_D^{}/M_N^{}\lesssim 6\times 10^{-6}$,  $m_D^2/M_N^{}\lesssim 4\times 10^{-5}\,\textrm{eV}$ for $M_N^{}\gtrsim 1\,\textrm{MeV}$. As a result, the mixing between the left- and right-handed neutrinos is negligible, while the neutrino mass matrix $m_\nu^{}$ is dominated by the type-II seesaw contribution $m_\nu^{\textrm{II}}$. Accordingly the right-handed neutrino mass matrix $M_N^{}$ can be well described by the left-handed neutrino mass matrix $m_\nu^{}$ up to an overall factor, i.e.
\begin{eqnarray}
m_\nu^{}=m_\nu^{\textrm{II}}=U_\nu^{\ast}\hat{m}_\nu^{}U_\nu^{\dagger}\,,
~~M_N^{}=\frac{\langle\Delta_R^{}\rangle}{\langle\Delta_L^{}\rangle}m_\nu^{}=U_\nu^{\ast}\hat{M}_N^{}U_\nu^{\dagger}\,.&&\nonumber\\
&&
\end{eqnarray}
The PMNS matrix is defined by $U=U_e^{\dagger}U_\nu^{}$ as usual. We can choose the base with the charged lepton mass
matrix $\tilde{m}_e^{}=\tilde{m}_e^\dagger$ being real and diagonal, and hence take $U_e^{}=1$ and $U_\nu^{}=U$.

If the mixing between the left- and right-handed charged gauge bosons $W_{L,R}^{\pm}$ is also small, one can compare quantitatively the left-handed and right-handed neutrinoless double beta decay processes. The $W_L$-$W_R$ mixing term will be induced only at one-loop level,
\begin{equation}
\mathcal{L}\supset \delta m_W^2 W_L^{-\mu}W_{R\mu}^{+}+\textrm{H.c.}
\end{equation}
with
\begin{equation}
\delta m_W^2\sim \frac{3g^2_{}}{32\pi^2_{}}m_t^{}m_b^{}\sim 3\,\textrm{GeV}^2_{}\,.
\end{equation}
For $M_{W_R^{}}^{}\gtrsim 1\,\textrm{TeV}$, the $W_L$-$W_R$ mixing will be extremely small, i.e. $\delta m_W^2/ M_{W_R^{}}^2\lesssim  3\times 10^{-6}_{}$.

\section{VI. Interpretation for the 750 GeV diphoton resonance at ATLAS and CMS}

\begin{figure}
\includegraphics[scale=0.4]{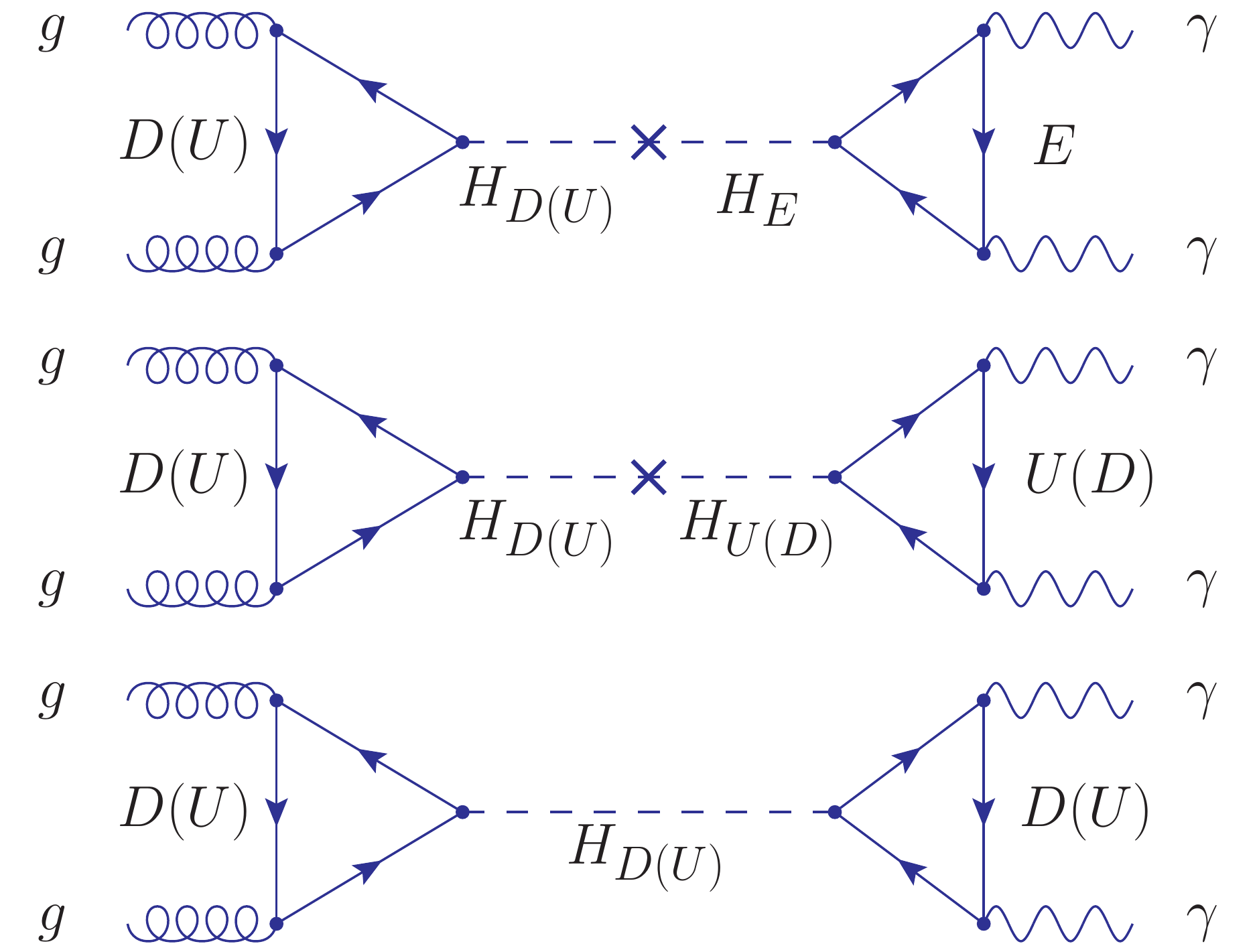}
\caption{The Feynman diagrams for $gg\to H\to \gamma\gamma$.  }
\label{diphoton}
\end{figure}

Recently, the ATLAS and CMS collaborations have reported a resonance of diphoton at the LHC Run-2 with energy
$\sqrt{s}=13$ TeV ~\cite{atlas, cms}. The cross section $\sigma(pp\to X \to\gamma\gamma)$ is roughly estimated
as $(6-10)$ fb, which $X$ dubbed as the  resonance.  In the model the CP-even singlet Higgs $H_{D,U,E}$ mix with each other and
one linear combination $H=c_{U}H_{U} + c_{D} H_{D} +c_{E} H_{E}$ could provide a good candidate for the $750$ GeV resonance. For simplicity, we consider a case when $H$ is a mixed state as
\begin{equation}
H=\cos\theta H_{U} + \sin\theta H_{E}.
\end{equation}
Using the narrow width approximation, one can write the cross section $\sigma(pp \to H\to \gamma\gamma)$ through the gluon-gluon fusion channel at the 13 TeV LHC as~\cite{Franceschini:2015kwy}
\begin{equation}
\sigma(pp\to H\to \gamma\gamma)=\frac{2137}{M_{H} \Gamma s} \Gamma(H\to gg)\Gamma(H\to \gamma\gamma)\,,
\end{equation}
where $M_{H}$ and $\Gamma$ are the mass and width of the scalar resonance, respectively.
The decay widths of $H\to \gamma\gamma$ and $H\to g g$ through the singlet fermion $(U, E)$ loops are given by~\cite{Djouadi:2005gi,Franceschini:2015kwy}
\begin{align}\label{decaywidth}
\frac{ \Gamma(H\to gg) }{M_{H}}&=\frac{\alpha_{s}^{2}}{2\pi^{3}} \cos^{2}\theta
\left(  \sum_{U} \frac{f_{U}}{2\sqrt{2}} \sqrt{r_{U}}\mathcal{I}(r_{U}) \right)^{2}\,,  \\
 \frac{\Gamma(H\to \gamma\gamma) }{M_{H}}&=\frac{\alpha_{em}^{2}}{16\pi^{3}} \left[
\frac{4\cos\theta}{3} \sum_{U} \frac{f_{U}}{\sqrt{2}} \sqrt{r_{U}}\mathcal{I}(r_{U})
 \right.   \nonumber\\
&+ \left.\sin\theta \sum_{E} \frac{f_{E}}{\sqrt{2}} \sqrt{r_{E}}\mathcal{I}(r_{E})
 \right]^{2}\,,
\end{align}
where $r_{{U,E}}=4M_{U,E}^{2}/M_{H}^{2}$ and
\begin{align}
\mathcal{I}(r)=1+(1-r)\arctan^{2}(1/\sqrt{r-1})\,.
\end{align}

\begin{figure}
\begin{center}
\includegraphics[width=0.4\textwidth]{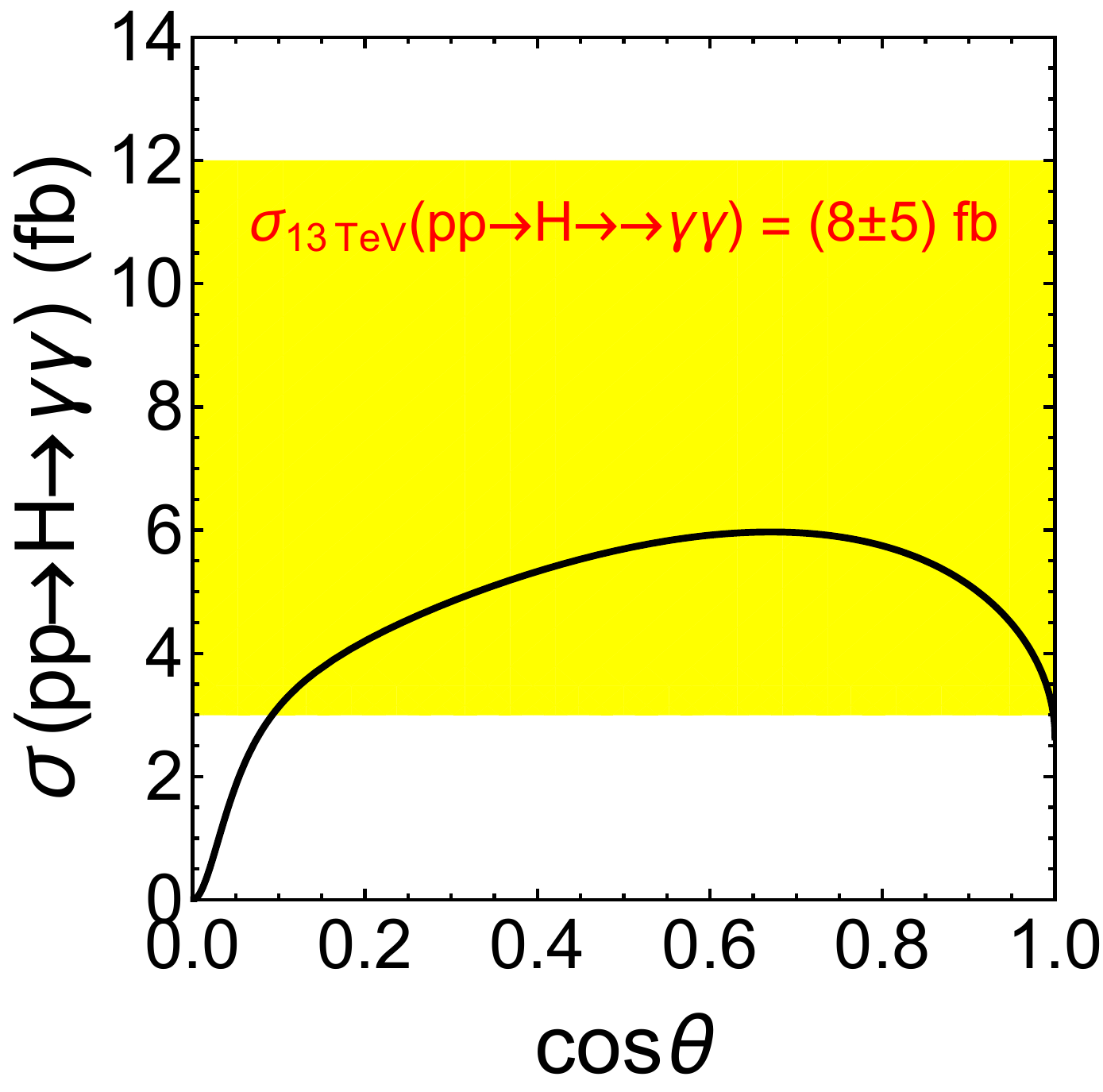}
\caption{The cross section $\sigma(pp\to H\to \gamma\gamma)$ as a function of the mixing $\cos\theta$. The yellow region
describes the range to explain the signal observed by the ATLAS and CMS collaborations. }
\label{fig3}
\end{center}
\end{figure}

For illustration we choose benchmark points $f_{U}=f_{E}=2.5, v_{U}=500 \text{GeV}, v_{E}=350 \text{GeV}$,
which predict the singlet fermion masses as $m_{U}=884~\text{GeV}$ and $m_{E}=619~\text{GeV}$.
Those vector-like quark masses satisfy the current constraints imposed by the ATLAS direct search~\cite{Aad:2015kqa}.
Assuming the total width is given by $\Gamma=\Gamma_{gg} + \Gamma_{\gamma\gamma}$,
we plot the cross section $\sigma(pp\to H\to\gamma\gamma)$ as a function of the mixing $\cos\theta$ in Fig.~\ref{fig3},
where $\alpha_{s}=0.1$ and $\alpha_{em}=1/128$ are used in our calculation. We also take into account of three generation vector-like quarks and leptons. The yellow region denotes the cross section needed to explain the diphoton excess.
The diphoton production rate can be fulfilled at a large range of parameter space, e.g. $0.1<\cos\theta<0.95$.
In our model the $\Gamma(H\to Z\gamma)\sim 0.6\Gamma(H\to \gamma\gamma)$, which is well consistent with current bound~\cite{Aad:2014fha}.

\section{VII. Conclusion}
In summary, an $SU(3)^{}_{c}\times SU(2)^{}_{L}\times SU(2)^{}_{R}\times U(1)^{}_{L}\times U(1)_{R}^{}\rightarrow SU(3)^{}_{c}\times SU(2)^{}_{L}\times SU(2)^{}_{R}\times U(1)^{}_{B-L}$ left-right symmetric framework is proposed for the universal seesaw scenario. Various vector-like $SU(2)$-singlet fermions and
Higgs scalars are introduced. The charged fermions obtain their masses through a ``seesaw'' way. The neutrino masses
are generated through the type-II seesaw with the help of $SU(2)$-triplet Higgs scalars, while the contribution from the canonical seesaw is negligible. The strong CP problem is solved by introducing a parity symmetry in the model. We also study the possibility of the singlet Higgs as a solution to the 750 GeV diphoton resonance observed by ATLAS and CMS at the LHC. We show that for reasonable parameters the singlet Higgs can be a good candidate for the resonance.

~\\
\textbf{Acknowledgement}:
The work is supported in part by the National Science Foundation of China (11175069, 11275009, 11422545)
and by the Shanghai Jiao Tong University under Grant No. WF220407201, the Recruitment Program for Young Professionals under Grant No. 15Z127060004 and the Shanghai Laboratory for Particle Physics and Cosmology under Grant No. 11DZ2260700.

\end{document}